# Energy-Efficient Cryogenic Neuromorphic Network with Superconducting Memristor

Md Mazharul Islam[1], Julia Steed[1], Karan Patel[1], Catherine Schuman[1], and Ahmedullah Aziz[1*]

[1] Department of EECS, University of Tennessee, Knoxville

*E-mail: aziz@utk.edu

**Abstract**

Cryogenic neuromorphic systems, inspired by the brain's unparalleled efficiency, present a promising paradigm for next-generation computing architectures. This work introduces a fully integrated neuromorphic framework that combines superconducting memristor (SM)-based spiking neurons and synapse topologies to achieve low-power neuromorphic network with non-volatile synaptic strength. This neurosynaptic framework is validated by implementing the cart-pole control task, a dynamic decision-making problem requiring real-time computation. Through detailed simulations, we demonstrate the network's ability to execute this task with an average fitness of ~5965 timesteps across 1000 randomized test episodes, with 40% achieving the target fitness of 15,000 timesteps (0.02s per timestep). The system achieves 23 distinct spiking rates across neurons, ensuring efficient information encoding. Our findings establish the potential of SM-based cryogenic neuromorphic systems to address the energy and scalability limitations of traditional computing, paving the way for biologically inspired, ultra-low-power computational frameworks.

## 1. Introduction

The persistent downsizing of transistors over the past decades has catalyzed remarkable technological progress, but this trend now faces fundamental physical and architectural bottlenecks[1]–[3]. The von Neumann bottleneck, characterized by the separation of memory and processing units, has exacerbated the challenges of handling the growing demands for data-intensive applications[4], [5]. To address these challenges, researchers have turned to alternative computational paradigms inspired by biological systems, particularly the human brain[6]. Neuromorphic computing, which seeks to emulate the brain's structure and functionality, has emerged as one of the most promising solutions[7]. Unlike conventional von Neumann architectures, neuromorphic systems integrate memory and computation into a single framework, enabling parallel processing and energy-efficient operation[8]. This paradigm has demonstrated significant potential for applications in artificial intelligence, machine learning, and real-time decision-making tasks[6], [9], [10].

However, despite these advantages, software-based implementations of neuromorphic computing fail to achieve the compactness, speed, and energy efficiency necessary for widespread adoption of neuromorhic nerwork[11]. High architectural complexity and significant area overhead

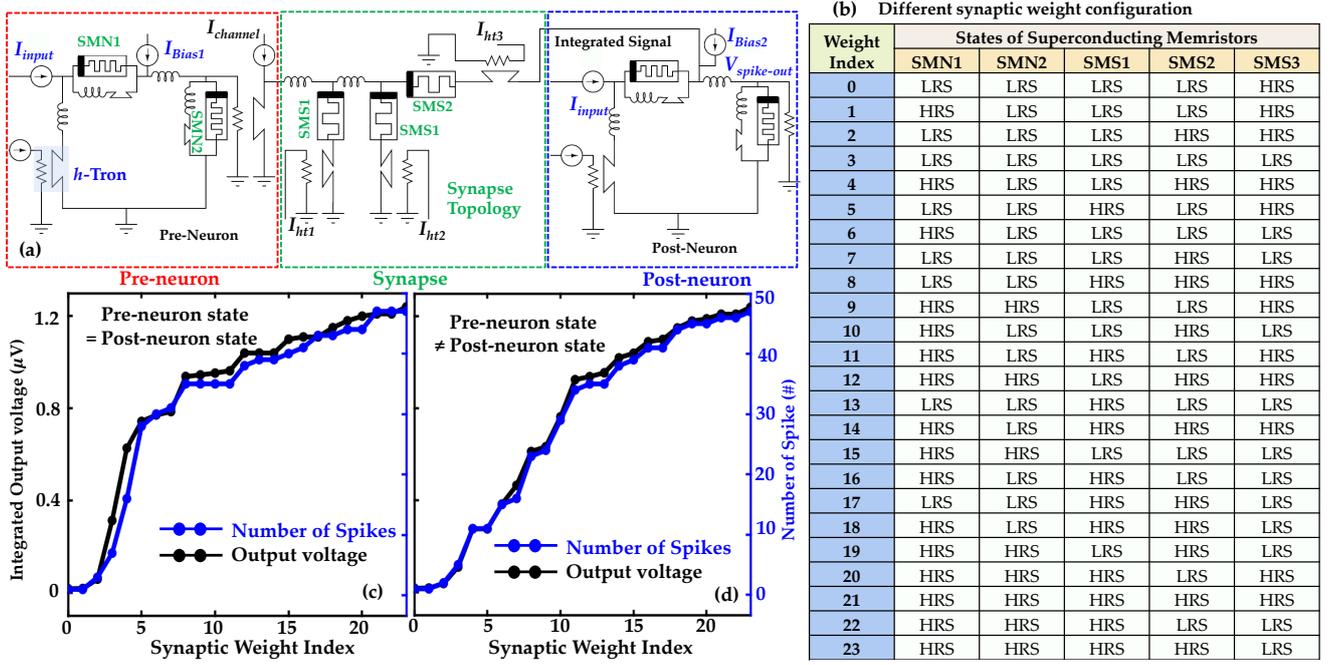

**Figure 1**: **(a)** Our proposed neurosynaptic framework where a synapse topology is placed between two neurons. **(b)** Corresponding SM configurations for 23 different synaptic weights. Generated post-neuron outputs for different synaptic weights when **(c)** pre-neuron state = post-neuron state, and **(d)** when pre-neuron state ≠ post-neuron state.

involved in their deployment pose further limitations in their implementation[12].

In response to these limitations, there has been a growing effort to develop hardware-based neuromorphic systems[8], [11], [13]. Pioneering implementations such as IBM's TrueNorth [14], [15] and Intel's Loihi [16] have demonstrated the feasibility of dedicated neuromorphic hardware. These systems offer substantial computational benefits over their software counterparts by leveraging specialized architectures optimized for spiking neural networks. However, despite these advancements, current neuromorphic hardware remains orders of magnitude less efficient than biological brains in terms of energy dissipation and functional density [7]. This disparity has driven a search for novel hardware paradigms that can more effectively mimic the energy-efficient and compact computational mechanisms of the brain.

Among emerging solutions, cryogenic neuromorphic systems have garnered significant attention for their unique capabilities [17]. Operating at cryogenic temperatures, these systems exploit the exceptional properties of superconducting materials, including ultra-low energy dissipation [18], high intrinsic switching speed [19], and dissipation-less interconnects [20]. These features make cryogenic neuromorphic systems an attractive platform for achieving energy-efficient, high-performance computing. Over the years, several superconducting devices, such as Josephson junctions, quantum phase slip junctions, and superconducting nanowires, have been proposed and employed in neuromorphic circuit designs. Early works, such as Crotty's Josephson junction-based spiking neuron [21] and Toomey's superconducting nanowire-based oscillatory neuron [19], [22], have laid a strong foundation for the development of cryogenic neuromorphic systems. However, these designs often suffer from practical limitations, including limited reconfigurability and precise biasing requirements.

Previously, these challenges were addressed by the introducion a dynamically reconfigurable spiking neuron topology that leverages superconducting memristors (SMs) and superconducting nanowires (SNWs) [23], [24]. This neuron topology achieved higher reconfigurability, offering non-volatile tunability with low energy dissipation and enhanced reconfigurability. Furthermore, potential of superconducting memristor-based synapse was also explored [25], focusing on its non-volatility and integrability in a diverse range of cryogenic hardware platforms. While both of these works provided critical insights into individual neuron and synapse components, the network-level functionality and integration of these elements remained unexplored.

This work aims to fill this gap by presenting a fully integrated cryogenic neuromorphic framework that combines the previously proposed SM-based neuron and synapse topology. By integrating these components into a cohesive network, we move beyond individual device-level demonstrations to evaluate their collective performance in executing a complete decision-making task. We apply this network to the classical pole-balancing problem [26], a standard test for neuromorphic systems that requires real-time decision-making and dynamic control. This task serves as a robust platform for assessing the feasibility and effectiveness of the integrated framework at the network level.

Through detailed simulation-based analyses, we examine the interactions between neurons and synapses, highlighting the critical parameters and design considerations that influence network performance. By evaluating the network's ability to perform pole balancing, we demonstrate the practical applicability of cryogenic neuromorphic systems for complex, real-world tasks. The findings underscore the potential of cryogenic neuromorphic systems to achieve unparalleled energy efficiency and scalability, paving the way for next-generation computing architectures.

The organization of this paper is as follows. Section 2 provides a brief overview of the previously proposed SM-based neuron and synapse topology. In section III, we discuss the complex interplay of the SM-based neuron and synapse topology and elaborate on the working principle of our proposed neurosynaptic network. Section IV describes the implementation of our proposed framework to perform a real-time pole-balancing task, along with a simulation-based analysis of the hardware implementation of this network.

## 2. Superconducting Memristor-based Neuron and Synapse Topology

### 2.1 Superconducting Spiking Neuron

The superconducting spiking neuron utilizes superconducting nanowires (SNWs) and superconducting memristors (SMs) to achieve spiking dynamics with non-volatile reconfigurability [23]. At its core lies a relaxation oscillator where an SNW is paired with an SM, allowing the modulation of oscillation frequencies by programming the resistive states of the SM. This topology introduces non-volatility, enabling four distinct spiking rates, derived from the SM's two resistive states. Additionally, the neuron can produce spikes with two different amplitudes, as demonstrated in previous studies. The incorporation of a heater cryotron (h-Tron) facilitates the independent programming of SM states, further enhancing design flexibility. Compared to earlier designs, this neuron achieves a 3.5× improvement in reconfigurability, making it a high-performance and energy-efficient component for cryogenic neuromorphic networks.

### 2.2 Superconducting Memristor-based Synapse

Superconducting Memristor-based synapse topology was previously proposed which offers non-volatile connection strength between two SM-based neurons previously discussed [25]. Similar to the conventional memristors, the SM exhibits a distinct pinched hysteresis loop in its current-voltage characteristics, enabling robust resistive state switching. The ability to achieve two distinct resistance levels makes the SM ideal for neuromorphic applications, where synaptic weights are key to network-level operation. For three different SMs, eight distinct synaptic weights are possibel in the SM-based synapse topology. The proposed topology integrates multiple superconducting devices to ensure seamless compatibility with spiking neurons across various superconducting platforms. This design underscores the scalability and energy efficiency of SM-based synapses, making them vital for advancing neuromorphic computing.

## 3. Simulation Framework and Device Parameters

For our simulation-based analysis, we have used HSPICE- an industry-standard cutting edge circuit simulator. For the SM, we have utilized physics-informed Verilog-A-based compact model. For the SNW, we have utilized a behavioral compact model previously developed. We have utlized previously reported prior analysis for choosing our device and material parameters for optimum performance. The device and material parameters are summarized in Table I. For training the Neuromorphic networrk for the Cart-pole Control task, EONS genetic algorithm [27] was used which was implemented within the TENNLab exploratory neuromorphic computing framework [10], [28].

## 4. Superconducting Neurosynaptic Framework

We utilize the SM-based neuron and synapse topologies to achieve a higher number of distinct synaptic states. A neuron augmented with a synapse topology incorporates a total of five superconducting memristors (SMs). Figure 1(a) illustrates the basic weight cell of our proposed neurosynaptic framework, where a synapse topology is positioned between two spiking elements.

The neuron topology was initially proposed by Crotty et al., utilizing Josephson junctions (JJs) as the oscillator elements. This design was later refined by Toomey et al., who replaced the JJs with superconducting nanowire (SNW)-based relaxation oscillators for improved fan-out capacity and efficiency. Islam *et al.* later modified the neuron topology by integrating superconducting memristors (SMs) and an h-Tron, as shown in Figure 1(a). This topology offers increased reconfigurability and non-volatility due to the distinct resistive states of the SMs. This design uses two coupled oscillators, where spikes are dynamically tuned via SM resistive states and fine-tuned with $I_{bias}$. The *h*-Tron enables independent SM programming, enhancing reconfigurability. The input current ($I_{in}$) switches the main oscillator's SNW, initiating coupled oscillations that generate spikes at the output, determined by the SM resistance and SNW inductance. These generated spikes can be integrated such that the resulting current reflects the number of spikes at the neuron's output. To achieve this functionality, the SM-based synapse topology is employed at the neuron output, as shown in Figure 1(a). This synapse topology consists of inductive nanowires connected to three superconducting memristors (SMs), each independently programmable via h-Tron devices. The h-Trons allow selective SM programming using a shared current ($I_{channel}$) for both programming and spike generation. Synaptic weights are updated by applying programming currents to specific SMs, adjusting their resistive states. During spike generation, incoming neuronal spikes trigger the *h*-Tron channel, directing current through the SM-SNW synapse structure. The integrated current at the synapse determines the strength of the input signal to the post-neuron. The integrated current from the synapse output is added to the bias current ($I_{Bias2}$) of the post-neuron. The integrated current is added to the post-neuron bias ($I_{Bias2}$), rather than the post-neuron input ($I_{in2}$), as the bias current has better tunability of the generated spikes at the output as reported in [24].

Table I: Materials and device simulation parameters

| Materials and Device Parameters | | | |
|---|---|---|---|
| **SNW parameters** | | **SM parameters** | |
| $R_{NW}$ | 5 kΩ | $R_{LRS}$ | 14.4 mΩ |
| $L_{NW}$ | 10 nH | $R_{HRS}$ | 98 mΩ |
| $I_c$, $I_r$ | 30 μA, 20 μA | $\gamma_0$ | 60° |

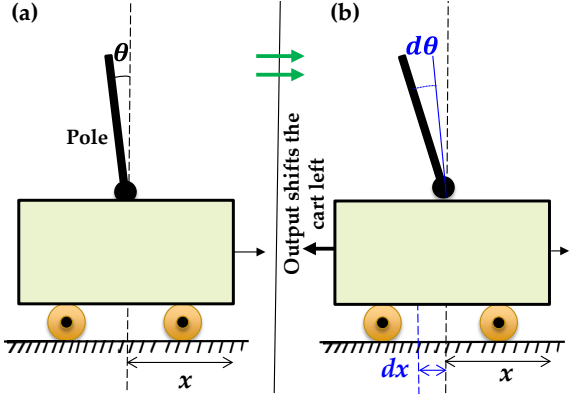

**Figure 2**: Cart-pole balancing control task. **(a)** Initial conditions showing pole position ($x$), pole-angle ($\theta$), **(b)** pole-angle velocity ($d\theta$), and pole velocity ($dx$). The corresponding output leads the cart to move to the left direction.

The integration of the post-synaptic current is influenced by the combination of SM resistances in both the pre-neuron and synapse. For five SMs (SMN1, SMN2, SMS1, SMS2, and SMS3), there are $2^5$=32 possible resistance combinations. However, due to some combinations producing indistinguishable integration rates, the system achieves 23 distinct spiking rates at the post-neuron output, as shown in Figure 1(b). For the subsequent analysis, these 23 unique synaptic weights are considered. Figures 1(c) and 1(d) depict the corresponding number of spikes generated at the post-neuron output over a 20 μs timeframe.

To evaluate the impact of the post-neuron's SM states, we considered two cases: when the post-neuron's SM states match the pre-neuron's SM states and when they differ. In both scenarios, the post-neuron output consistently produces 23 distinct spiking rates. For simplicity, we are considering the first case for our subsequent analysis where the post-neuron's SM states match the pre-neuron's SM states.

## 5. Cart-pole Control Task Using the Proposed Neurosynaptic Framework

Control applications are a key domain for machine learning and neural networks, presenting unique challenges due to the inherent time dependency. Unlike static tasks like image classification, control tasks involve dynamic interactions where the agent's actions influence subsequent observations dynamically, making traditional backpropagation inadequate for training. Neuromorphic processors, with spiking neural networks (SNNs) that inherently integrate time and co-locate memory with processing, are well-suited for such tasks. However, designing or training of SNNs remains a significant challenge. SNNs have shown success in various control problems, particularly where hardware constraints like size, weight, power, and latency are critical.

The "cart-pole" problem is an example of a classic control task (Fig.2). Here, a wheeled cart on a one-dimensional track must balance a pole while staying within track bounds, with actions taken at regular intervals to maintain stability. The problem is a popular benchmark due to its simplicity and accessibility. Agents receive four observations—cart position ($x$), cart velocity ($dx$), pole angle ($\theta$), and pole angular velocity ($d\theta$) (Fig. 2(a))—and can take one of two actions at each step: push left or push right (Fig. 2(b)). In this study, we address the standard cart-pole problem with a mission duration of up to 15,000 timesteps (0.02s per timestep), averaged across 1,000 random initial positions (i.e., 1,000 episodes). Training is performed using the EONS genetic algorithm [27], implemented within the TENNLab exploratory neuromorphic computing framework [10], [28]. Input values are encoded as spike rates in the input neurons [26], while outputs are then decoded using a voting

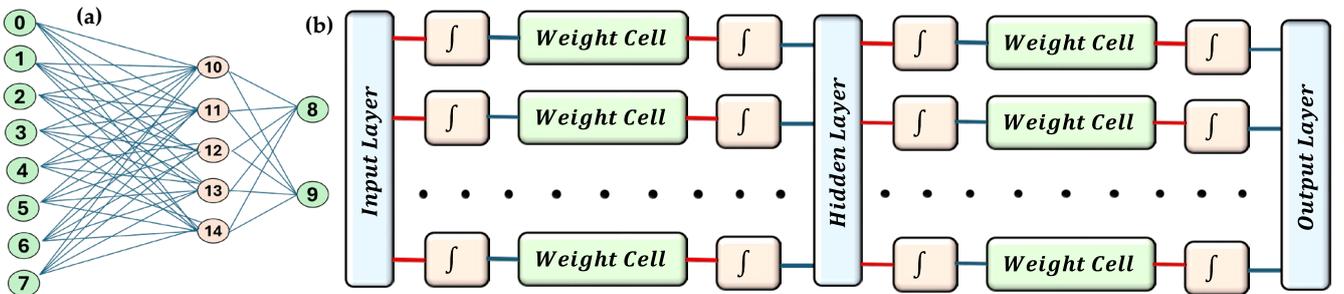

**Figure 3**: **(a)** Neuromorphic network connection for Cart-pole control task. **(b)** General Configuration of our implemented neurosynaptic network.

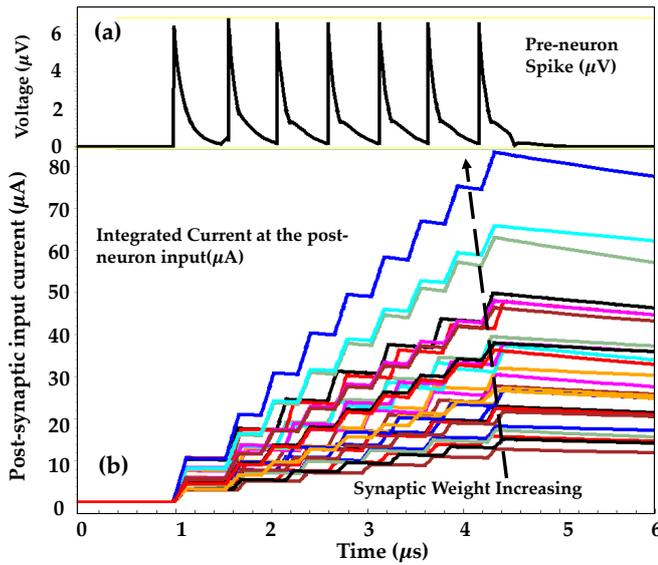

**Figure 4**: **(a)** Output spike of the input neurons in the input layer. **(b)** Integrated Current that is fed in the hidden layer input

mechanism, where the neuron with the highest spike count determines the action.

The optimized network for the task is depicted in Fig. 3(a). Our goal is to implement the network parameters, which are learned off-chip, within the proposed neurosynaptic framework shown in Fig. 1(a). The neuron model used is based on the SM-based neuron topology discussed in section 2.2. For implementing the inter neuron weights, we utilize the weight cell illustrated in Fig. 1(a), which supports 23 distinct weight values as demonstrated in Figs. 1(b-d). The general structure for implementing the network within our proposed framework is summarized in Fig. 3(b). In the input layer, each neuron generates a specific number of spikes, which are integrated and fed into the weight cell (Fig. 4). The output spike generated by the weight cell is determined by both the input strength and the stored weight (Fig.4(b)). The resulting spikes are then integrated and passed as input to the hidden layer (Fig.3). This approach repeats from the hidden layer to the output layer. For the cart-pole control task, the network has two output neurons, each corresponding to a shift direction: left or right.

The network implementation within our framework closely aligns with the results of the off-chip trained network, as illustrated in Fig. 5. Specifically, the spikes generated at the output neurons in our proposed neurosynaptic network closely resemble and are proportional to the outputs trained using the EONs algorithm (Figs. 5(e-f,h-i)). With an appropriate decoder circuit at the output, these results can be accurately interpreted. Finally, Fig. 6 presents a histogram summarizing the fitness of our network, which was evaluated 1000 random episodes, each starting from a different initial conditions. ~40% of total episodes achieve the target fitness of 15000 timesteps. The average fitness of 1000 test episodes is ~5965 timesteps.

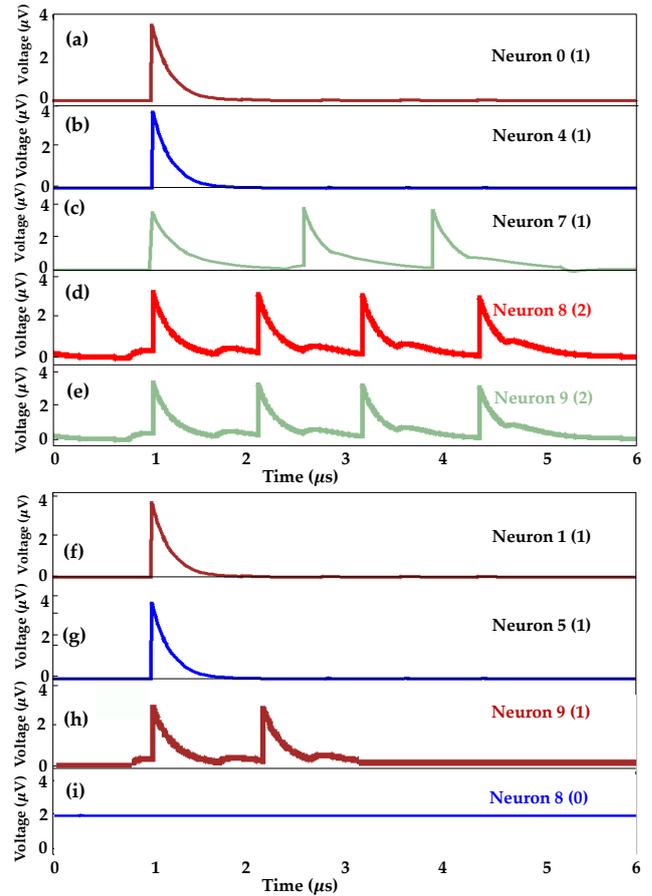

**Figure 5**: Input and output spikes of our implemented neuromorphic network for the cart-pole balancing control task. Two different combinations of results are illustrated (Figs(a-e), Figs(f-i)). Trained values are written inside the bracket. Spike output in input neurons: **(a)** Neuron 0, **(b)** Neuron 4, and **(c)** Neuron 7, and corresponding output neurons: **(d)** Neuron 8, **(e)** Neuron 9. Spike output in input neurons: **(f)** Neuron 1, **(g)** Neuron 5, and corresponding output neurons: **(h)** Neuron 9, **(i)** Neuron 8. The number of generated spikes is proportional to the theoretical test result.

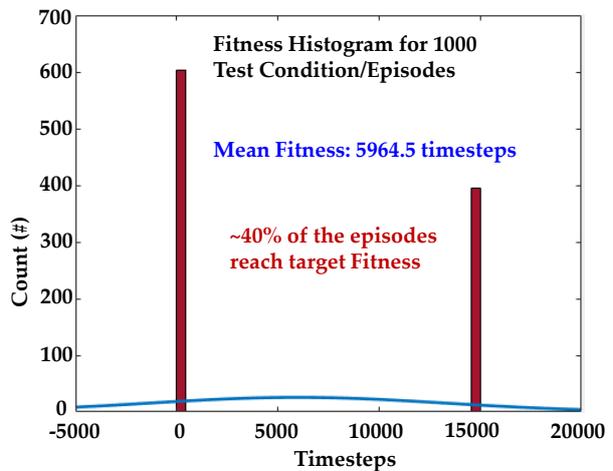

**Figure 6**: Fitness result of the Cart-pole control tasks for 1000-episodes.

## 6. Conclusion

This work presents a novel cryogenic neuromorphic framework that integrates superconducting memristor-based neuron and synapse topologies into a scalable, energy-efficient network. Through extensive simulations, we demonstrate the framework's ability to tackle real-world tasks such as the cart-pole control problem, achieving promising results in terms of targeted fitness. The insights gained from this study reveal the critical parameters influencing neurosynaptic network performance and establish a foundation for future developments in superconducting neuromorphic systems. This research paves the way for leveraging cryogenic technologies in building compact, high-speed, and energy-efficient computing systems, bringing us closer to mimicking the unparalleled efficiency of the human brain.